\documentclass[aps,pra,reprint]{revtex4-2}

\usepackage{amsmath,amssymb,bm}
\usepackage{graphicx}
\usepackage{hyperref}
\newcommand{\avg}[1]{\left\langle #1 \right\rangle}
\newcommand{\ii}{\mathrm{i}}
\newcommand{\dd}{\mathrm{d}}

\begin{document}

\title{Effective-frequency shear organizes chaotic response under higher-order coupling}

\author{Kaiming Luo}
\email{kmluo24@m.fudan.edu.cn}
\affiliation{College of Future Information Technology, Fudan University, Shanghai 200438, China}

\begin{abstract}
How higher-order interactions reorganize chaotic responses beyond established coherence-based descriptions remains incompletely resolved. In amplitude-active systems, such responses are often interpreted through phase coherence, although coherence alone does not determine collective instability. Here we study a minimal globally coupled quartet of nonisochronous Stuart--Landau oscillators with pairwise and symmetric three-body interactions. Because the pairwise system already supports a connected chaotic branch, the higher-order channel reconstructs rather than creates irregular dynamics. We show that this reconstruction is organized more directly by effective-frequency shear than by phase coherence. Higher-order coupling regulates amplitude heterogeneity, nonisochronicity converts it into differential angular velocities, and the resulting shear structures the Lyapunov response. This reduced shear-based description reveals an indirect amplitude--shear pathway for higher-order control of chaotic dynamics.
\end{abstract}

\maketitle
\textbf{This work identifies a novel mechanism by which higher-order interactions regulate chaotic dynamics in nonisochronous oscillator networks. Unlike conventional analyses that focus on phase coherence, we show that effective-frequency shear—arising from amplitude heterogeneity and nonisochronicity—is the primary organizing variable controlling instability. Using a minimal four-oscillator system and finite-size extensions, we demonstrate a structured pathway K$_3 \rightarrow \sigma_r \rightarrow \sigma_\Omega \rightarrow \lambda_\mathrm{max}$, providing a clear causal hierarchy. This mechanism highlights the hidden internal degrees of freedom as the key mediators of chaos and offers a general framework for understanding higher-order coupling effects in complex nonlinear systems.}

\section{Introduction}

Higher-order interactions can reorganize attractor stability rather than merely shift pairwise thresholds \cite{battiston20,battiston21phys,zhang23}. Recent studies have demonstrated their role in neuronal synchrony, multilayer synchronization, swarmalator dynamics, and neuron models with nonpairwise coupling \cite{majhi22review,majhi25neuronal,pal24global,anwar24swarmalators,parastesh22hindmarsh}. In nonlinear oscillator systems, many-body channels can reshape branch persistence and redistribute fluctuations across dynamical sectors. Here we address a more specific question: whether higher-order coupling regulates the stability of a preexisting chaotic collective state, and which dynamical quantity mediates this control.

This question is easily obscured when the dynamics are described only through coarse coherence indicators. A phase order parameter captures average alignment but does not determine whether instability is governed by phase locking, amplitude disorder, or differential angular velocities. States with similar coherence can nevertheless differ strongly in internal frequency spread and Lyapunov growth. For higher-order control of chaos, the key distinction is therefore whether a many-body term acts directly on synchrony or indirectly through a hidden internal degree of freedom.

Stuart–Landau oscillators provide a natural setting in which to resolve this issue because amplitude remains dynamical rather than slaved \cite{aronson90,matthews91,aranson02world}. In nonisochronous ensembles, radial fluctuations feed back into angular dynamics through amplitude-dependent rotation rates \cite{brown04phase,nakao16,pietras19,luo2024effects}. This amplitude-to-phase conversion channel is where shear emerges: once amplitude heterogeneity survives, it is converted into differential angular velocities that generate the stretching and reinjection processes required for chaotic growth.

Coherence and shear must therefore be distinguished. Coherence compresses the collective state onto a single alignment scale, whereas shear measures the spread of effective angular velocities after amplitude disorder has been converted into phase-speed disorder. When the goal is to explain why chaos appears or disappears, shear lies closer to the instability mechanism. A mechanism-based account of higher-order control must therefore identify how amplitude heterogeneity is regulated, how it is converted into shear, and how that shear determines the instability response.

Earlier work contains elements of this picture but not its full causal closure. Globally coupled amplitude oscillators can exhibit collective irregularity and complex mean-field dynamics even in symmetric settings \cite{hakim92,nakagawa94,rosenblum07}. Related Stuart–Landau studies have revealed nonlinear collective states and irregular spectra pointing to an essential amplitude sector \cite{schmidt12exp,luccioli19,clustersing19}, while delay-induced irregularity reinforces the same mechanism \cite{delaycgl19}. Higher-order interaction studies have further shown that nonpairwise couplings reorganize collective branches and stability boundaries \cite{gambuzza21stable,gallo22,multipop22,nonlocal23}. However, higher-order Stuart–Landau models remain relatively scarce, and even when higher-order coupling visibly reshapes the dynamics, the controlling variable of the chaotic response remains unclear \cite{dutta23sd,verma24tip}.

The difficulty is not the lack of diagnostics. Lyapunov exponents and tangent-space methods reliably identify chaotic regimes, and shear-based reasoning provides a direct route to stretching-based interpretations \cite{benettin80a,benettin80b,wolf85lyap,lin08shear}. What is missing is a mechanism-based hierarchy that separates a coarse coherence marker from an upstream mediator and a controlling variable from the final instability output. Without such a hierarchy, statements about nonmonotonic chaos remain ambiguous, because onset location, chaotic window width, peak instability, and onset sharpness need not respond identically to higher-order coupling. Earlier onset does not necessarily imply stronger chaos, and weaker peak instability does not imply immediate loss of chaotic support.

In this work, we study a fully connected quartet of Stuart–Landau oscillators with pairwise coupling, symmetric three-body coupling, heterogeneous intrinsic frequencies, and frequency-dependent nonisochronicity. This minimal setting isolates the competition among amplitude heterogeneity, nonisochronicity, and higher-order feedback in the smallest globally coupled system that supports collective chaos. The pairwise baseline already contains a connected chaotic window, so higher-order coupling acts on a preexisting irregular branch.

The three-body channel reconstructs this branch indirectly: it regulates amplitude dispersion, nonisochronicity converts this dispersion into effective-frequency shear, and the resulting shear governs the instability more directly than coherence. Finite-size extensions to larger fully connected systems show that the same amplitude–shear route remains operative, with its quantitative strength modulated by system size. The quartet thus provides a minimal setting in which this mechanism can be identified clearly.

The paper is organized around this causal hierarchy. Section II introduces the model, defines the observable hierarchy, and establishes the pairwise baseline as the preexisting chaotic branch. Section III tracks how the three-body channel reshapes onset location, chaotic support, peak instability, and onset sharpness without treating these diagnostics as equivalent. Section IV closes the amplitude–shear mechanism through amplitude–phase reduction, reduced collapse, control slices, linear-stability projection, and finite-size robustness. The final sections discuss the broader implication that higher-order interactions regulate hidden internal degrees of freedom and summarize the resulting indirect control of chaos.

\begin{figure*}[t]
    \includegraphics[width=0.9\linewidth]{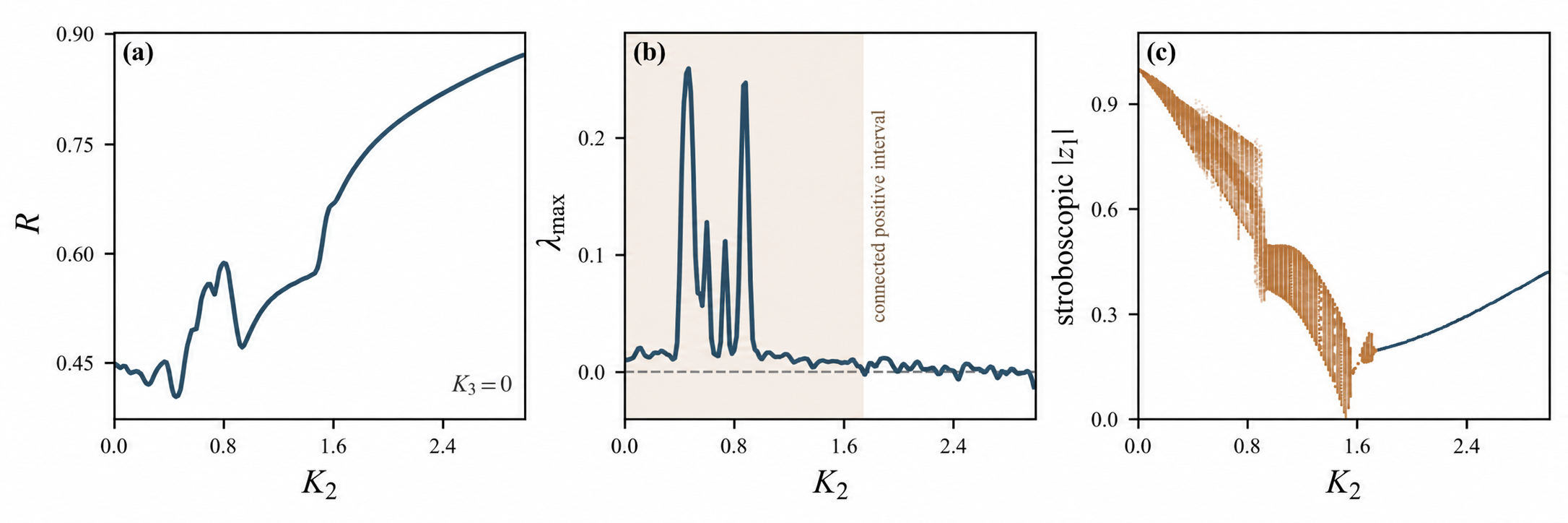}
\caption{Baseline diagnostics of the pairwise quartet at \(K_3=0\). (a) Time-averaged order parameter \(R\). (b) Largest Lyapunov exponent \(\lambda_{\max}\), with the shaded region marking the connected chaotic window. (c) Stroboscopic values of \(|z_1|\), showing the attractor-geometry signature of the same branch. The positive Lyapunov interval and the corresponding irregular stroboscopic structure show that chaos already exists in the pairwise system, so higher-order coupling reshapes a preexisting chaotic branch rather than creating chaos from a regular state.}
    \label{fig:baseline}
\end{figure*}

\section{Model, Observables, and Pairwise Baseline}

The dynamics are governed by the network Stuart-Landau equation for complex states $z_i=r_i e^{\ii\theta_i}$ \cite{aronson90,matthews91,aranson02world,garcia12}:
\begin{equation}
\begin{aligned}
\dot{z}_i ={}&
\left(1+\ii\omega_i-(1+\ii c_i)|z_i|^2\right)z_i \\
&+\frac{K_2}{N}\sum_{j=1}^{N} A_{ij}(z_j-z_i) \\
&+\frac{K_3}{N^2}\sum_{j,k=1}^{N} B_{ijk}(z_j z_k^\ast-z_i),
\qquad i=1,\dots,N.
\end{aligned}
\label{eq:full_model}
\end{equation}
The pairwise-plus-triadic Stuart-Landau form in Eq.~(\ref{eq:full_model}) follows the same higher-order Stuart-Landau modeling direction adopted in recent studies of triadic and repulsive higher-order couplings \cite{dutta23sd,verma24tip}. Here $\omega_i$ are the intrinsic frequencies, $c_i$ are the nonisochronicity coefficients, $K_2$ is the pairwise coupling strength, $K_3$ is the symmetric three-body coupling strength, $A_{ij}$ is the pairwise adjacency matrix, and $B_{ijk}$ is the symmetric three-body connectivity tensor. This system is minimal yet dynamically complete for the present problem because it retains every ingredient required by the mechanism in a single equation: local amplitude dynamics, oscillator mismatch, nonisochronous amplitude-to-phase conversion, pairwise homogenization, and higher-order feedback. Removing any of these ingredients would simplify the model, but it would also destroy one step of the causal chain that we want to resolve.

We specialize to the fully connected case with $N=4$, so that $A_{ij}=1$ and $B_{ijk}=1$ for all indices. The point of the quartet is not topological complexity.  It is the smallest globally coupled setting in which amplitude heterogeneity,
frequency-dependent nonisochronicity, and higher-order regulation can still compete
strongly enough to support collective irregularity.
The quartet also provides a nondegenerate fully connected higher-order structure,
in which each oscillator is embedded in multiple triadic feedback channels and
the three-body term can act on a genuinely collective amplitude field.
In that sense the quartet is a minimal mechanism carrier. It is small enough that the geometry of the branch can still be read directly, yet large enough that the amplitude field is genuinely collective and not reducible to a single oscillator correction.

The intrinsic frequencies are taken as $\omega_i=\omega+\Delta\omega\,\xi_i$ with $\omega=2.0$, $\Delta\omega=0.5$, and $\xi_i=[-1,-1/3,1/3,1]$. The nonisochronicity coefficients are chosen as $c_i=k\omega_i$ with $k=-5$ unless stated otherwise. This proportional choice is important because it makes the amplitude-to-phase conversion frequency dependent. Oscillators with different natural frequencies do not merely rotate at different base rates; they also convert amplitude fluctuations into angular-velocity fluctuations with different efficiencies. That structure is what later allows amplitude disorder to become effective-frequency shear.

The fully connected setting removes a second source of interpretation that would otherwise obscure the mechanism, namely spatial organization. There are no motifs, sparse neighborhoods, or long-range shortcuts that could independently generate inhomogeneous local environments. Any persistent heterogeneity must therefore arise from the internal amplitude dynamics, from the imposed frequency mismatch, or from the competition between pairwise and higher-order mean-field feedback. This is precisely the regime in which one can ask, without topological ambiguity, whether higher-order coupling regulates coherence, amplitude dispersion, effective-frequency shear, or the final instability itself.

For the fully connected quartet the mean field is
\begin{equation}
Z=\frac{1}{N}\sum_{j=1}^{N} z_j,
\label{eq:mean_field}
\end{equation}
and the identity
\begin{equation}
\begin{aligned}
\frac{1}{N^2}\sum_{j,k=1}^{N} z_j z_k^\ast
={}&
\left(\frac{1}{N}\sum_{j=1}^{N} z_j\right)
\left(\frac{1}{N}\sum_{k=1}^{N} z_k\right)^\ast \\
={}& |Z|^2
\end{aligned}
\label{eq:meanfield_identity}
\end{equation}
Using Eq.~(\ref{eq:mean_field}) and Eq.~(\ref{eq:meanfield_identity}), Eq.~(\ref{eq:full_model}) reduces to
\begin{equation}
\begin{aligned}
\dot{z}_i ={}&
\left(1+\ii\omega_i-(1+\ii c_i)|z_i|^2\right)z_i \\
&+K_2(Z-z_i)+K_3(|Z|^2-z_i).
\end{aligned}
\label{eq:compact_model}
\end{equation}
The reduced form in Eq.~(\ref{eq:compact_model}) is used in the scans below because it keeps the global structure transparent while preserving the full amplitude--phase dynamics.

For each parameter point we follow the long-time bounded dynamics of Eq.~(\ref{eq:compact_model}) after transients have decayed, evaluate time averages on the resulting attractor, and compute $\lambda_{\max}$ from the tangent-space evolution with standard reorthonormalization \cite{benettin80a,benettin80b,wolf85lyap}. The numerical protocol is therefore aligned with the logic of the paper: the trajectory determines the attractor, the Lyapunov calculation determines whether that attractor is regular or chaotic, and the time-averaged observables diagnose which internal sector of the dynamics is being regulated.

This ordering of diagnostics is important because it prevents the interpretation from becoming circular. We do not infer chaos from a coherence measure and then use the Lyapunov exponent only as a secondary confirmation. Instead, the Lyapunov exponent defines the instability output, while the remaining observables are asked to explain why that output changes. The observable hierarchy is therefore built to support a mechanism, not merely to provide a descriptive bundle of curves.

The observables are not introduced as four equally important diagnostics because they do not play equivalent roles in the mechanism. The final instability output is the largest Lyapunov exponent $\lambda_{\max}$, which measures the asymptotic exponential separation of nearby trajectories and therefore records whether the collective state is dynamically unstable. As a coarse coherence indicator we use
\begin{equation}
R=\avg{\left|\frac{1}{N}\sum_{j=1}^{N} e^{\ii\theta_j}\right|}_t,
\label{eq:R}
\end{equation}
which measures phase alignment across the quartet \cite{kuramoto75,acebron05,rodrigues16}. The role of $R$ is deliberately limited from the outset. It is useful for showing that the dynamics remain only partially coherent, but it cannot distinguish whether the surviving instability is controlled by a phase-locking deficit, by amplitude disorder, or by shear. It is therefore an auxiliary observable rather than a controlling one.

The upstream mediator is the amplitude dispersion
\begin{equation}
\sigma_r=
\avg{\left[\frac{1}{N}\sum_{j=1}^{N}(r_j-\bar r)^2\right]^{1/2}}_t,
\qquad
\bar r=\frac{1}{N}\sum_{j=1}^{N} r_j,
\label{eq:sigma_r}
\end{equation}
which measures amplitude heterogeneity across the oscillators. The effective-frequency dispersion,
\begin{equation}
\sigma_\Omega=
\avg{\left[\frac{1}{N}\sum_{j=1}^{N}(\Omega_j-\bar\Omega)^2\right]^{1/2}}_t,
\qquad
\bar\Omega=\frac{1}{N}\sum_{j=1}^{N}\Omega_j,
\label{eq:sigma_omega}
\end{equation}
with $\Omega_i=\operatorname{Im}(\dot z_i/z_i)$. The distinction between $\sigma_r$ and $\sigma_\Omega$ is the central observable hierarchy of the paper. $\sigma_r$ tells us whether amplitude heterogeneity survives, whereas $\sigma_\Omega$ tells us whether that heterogeneity has been converted into differential angular velocities. Among these observables, only $\sigma_\Omega$ directly measures the differential angular velocities responsible for shear-driven stretching. The hierarchy maintained throughout the paper is therefore $R < \sigma_r < \sigma_\Omega < \lambda_{\max}$ in causal significance: $R$ is coarse, $\sigma_r$ is upstream, $\sigma_\Omega$ is the controlling variable, and $\lambda_{\max}$ is the final instability output.

It is useful to emphasize what $\sigma_\Omega$ does and does not represent. It is not a static measure of quenched frequency mismatch, because the imposed frequency heterogeneity is fixed throughout the study. Nor is it reducible to the spread of coupling terms alone, because the quantity is evaluated on the actual nonlinear trajectory and therefore includes the amplitude-dependent conversion generated by nonisochronicity. In other words, $\sigma_\Omega$ is a dynamical shear measure, not a parameter-level disorder measure. That is why it serves as the controlling variable even when the bare mismatch is unchanged.

The pairwise limit $K_3=0$ then provides the reference object on which the whole argument depends. If the baseline state were regular throughout the scan, the natural conclusion would be that higher-order coupling creates chaos. That is not what happens here. Fig.~\ref{fig:baseline} shows that the pairwise quartet already supports a connected chaotic window, so the three-body channel acts on an already existing irregular branch. This baseline is therefore not a preliminary comparison but the dynamical object that will later be reconstructed.

The three panels of Fig.~\ref{fig:baseline} serve different diagnostic roles. Panel (a) shows the time-averaged coherence measure $R$ and establishes that the baseline dynamics remain only partially coherent; it is useful as a coarse orientation, but it does not identify the control variable. Panel (b) is the core baseline plot because it isolates a connected interval of positive $\lambda_{\max}$ and therefore defines the preexisting chaotic window directly in the instability output. Panel (c) records the attractor geometry of the same branch through the spread of the stroboscopic amplitude cloud. The geometric broadening and the positive Lyapunov exponent point to the same object from two different sides: one dynamical, one geometric.

That baseline object is already informative about the later mechanism. Because the chaotic branch is visible in both the Lyapunov scan and the stroboscopic amplitude cloud, it is clear that the irregularity is not a purely phase-based effect. The branch has a resolved amplitude sector even before the higher-order channel is turned on. This makes the later emphasis on amplitude dispersion natural rather than retrospective: the three-body term will act on a branch whose geometry already contains the amplitude fluctuations needed for a shear-based route to instability.

This distinction is essential for the interpretation of higher-order coupling. Because the pairwise system already contains the irregular branch, later changes in onset, width, or instability cannot be read as the birth of a wholly new chaotic state. They must instead be read as the reconstruction of an existing route to chaos. Consequently, the main question of the paper is not whether the three-body channel enhances or suppresses chaos in a coarse sense, but what quantity it actually regulates.

\begin{figure*}[t]
\centering
\includegraphics[width=1\linewidth]{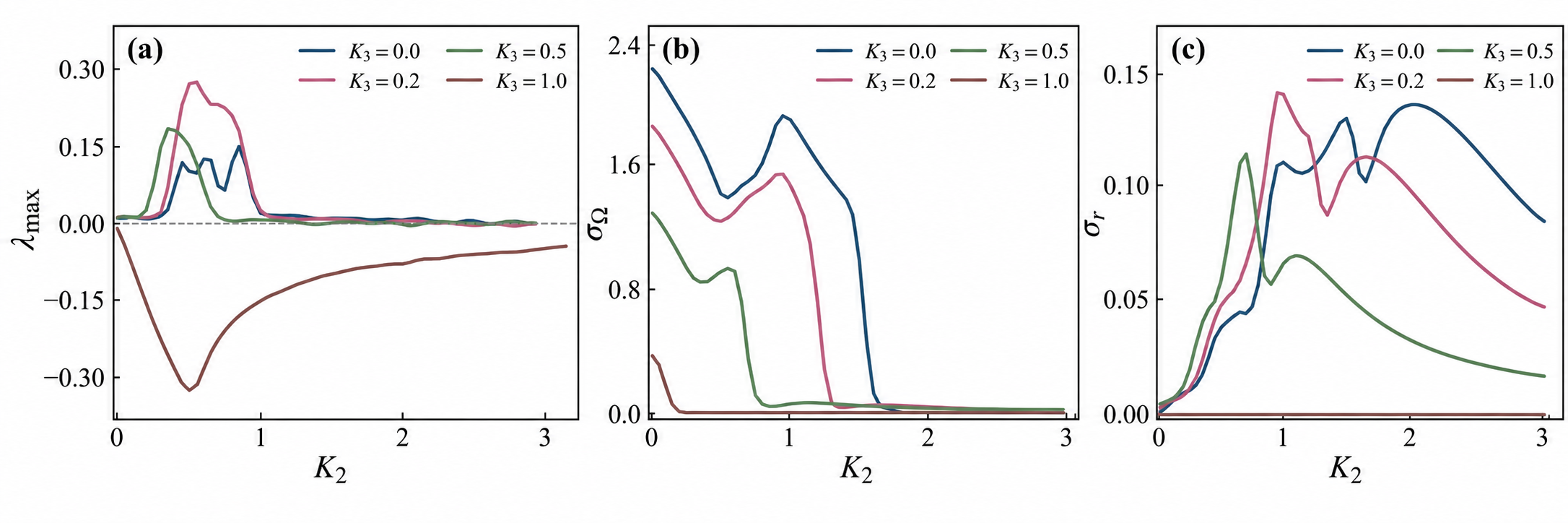}
\caption{Representative tracking of the preexisting chaotic branch as the pairwise coupling \(K_2\) is scanned for several fixed values of the triadic coupling \(K_3\). 
(a) The largest Lyapunov exponent \(\lambda_{\max}\) locates the connected chaotic window; the dashed line denotes \(\lambda_{\max}=0\), and positive values indicate chaos. 
The branch already exists at \(K_3=0\), so the higher-order interaction reshapes rather than creates the chaotic response. 
(b) The effective-frequency shear \(\sigma_\Omega\) tracks the shear sector of the same branch. 
(c) The amplitude dispersion \(\sigma_r\) tracks the upstream amplitude sector. 
Together, panels (b) and (c) show that increasing \(K_3\) first regulates amplitude disorder, which then reduces effective-frequency shear and ultimately narrows or suppresses the Lyapunov response.}
\label{fig:route}
\end{figure*}

\section{Nonmonotonic Reconstruction of the Chaotic Window}

Once the pairwise baseline is identified, the effect of higher-order coupling can be posed as a tracking problem for one and the same branch. Fig.~\ref{fig:route} should therefore not be read as three unrelated scans. It follows the same chaotic window as $K_3$ is increased and asks how the instability, shear, and amplitude sectors are reconstructed. In this formulation the three-body term does not open a detached chaotic branch elsewhere in parameter space. Instead, it reshapes the support, intensity, and geometric expression of the branch already present at $K_3=0$.

The three panels of Fig.~\ref{fig:route} already suggest why a single coherence language is insufficient. Panel (a) tracks the instability output through $\lambda_{\max}$ and therefore identifies where the connected chaotic window persists. Panel (b) tracks the same branch through $\sigma_\Omega$, while panel (c) follows the upstream amplitude sector through $\sigma_r$. Taken together, they show that the branch is not rigidly translated by $K_3$; rather, its amplitude content, its shear content, and its Lyapunov response change at different rates. This is the first indication that higher-order coupling acts upstream of the instability output.

The reconstruction is layered. Weak $K_3$ shifts the onset to smaller $K_2$, broadens the connected chaotic support, and increases the peak instability. In this regime the three-body channel enlarges the accessible amplitude disorder and thereby improves the efficiency with which nonisochronicity generates shear. At intermediate $K_3$, the onset may continue to drift left, but the positive-Lyapunov support begins to contract and the instability weakens. Strong $K_3$ eventually collapses the window altogether. The same coupling parameter therefore produces three qualitatively distinct regimes: early enhancement, partial weakening, and final suppression.

This layered response is already mirrored by the auxiliary scans in Figs.~\ref{fig:route}(b) and \ref{fig:route}(c). In the weak-coupling regime, the rise of $\sigma_r$ is accompanied by a rise of $\sigma_\Omega$, indicating that the enhanced chaotic support is not an isolated Lyapunov phenomenon but part of a broader amplification of the amplitude and shear sectors. At intermediate and strong $K_3$, the two quantities no longer grow in tandem with the leftward onset shift, which is the first direct hint that onset location is not the correct proxy for instability strength. The branch survives earlier in parameter space even while the internal disorder that feeds it is being reduced.

Earlier onset does not necessarily imply stronger chaos. This point is crucial because onset location and instability strength are different diagnostics of the same reconstructed branch. A leftward drift of the onset only means that the branch becomes chaotic at smaller $K_2$; it does not imply that the branch occupies more parameter space or that its strongest stretching rate has increased. That distinction is already visible in Fig.~\ref{fig:route}: the onset may move left even while the interval of positive $\lambda_{\max}$ narrows and the peak Lyapunov response drops. The higher-order channel therefore does not simply ``advance chaos'' or ``delay chaos.'' It changes several nonequivalent properties of the same branch.

\begin{figure*}[t]
\centering
\includegraphics[width=0.8\linewidth]{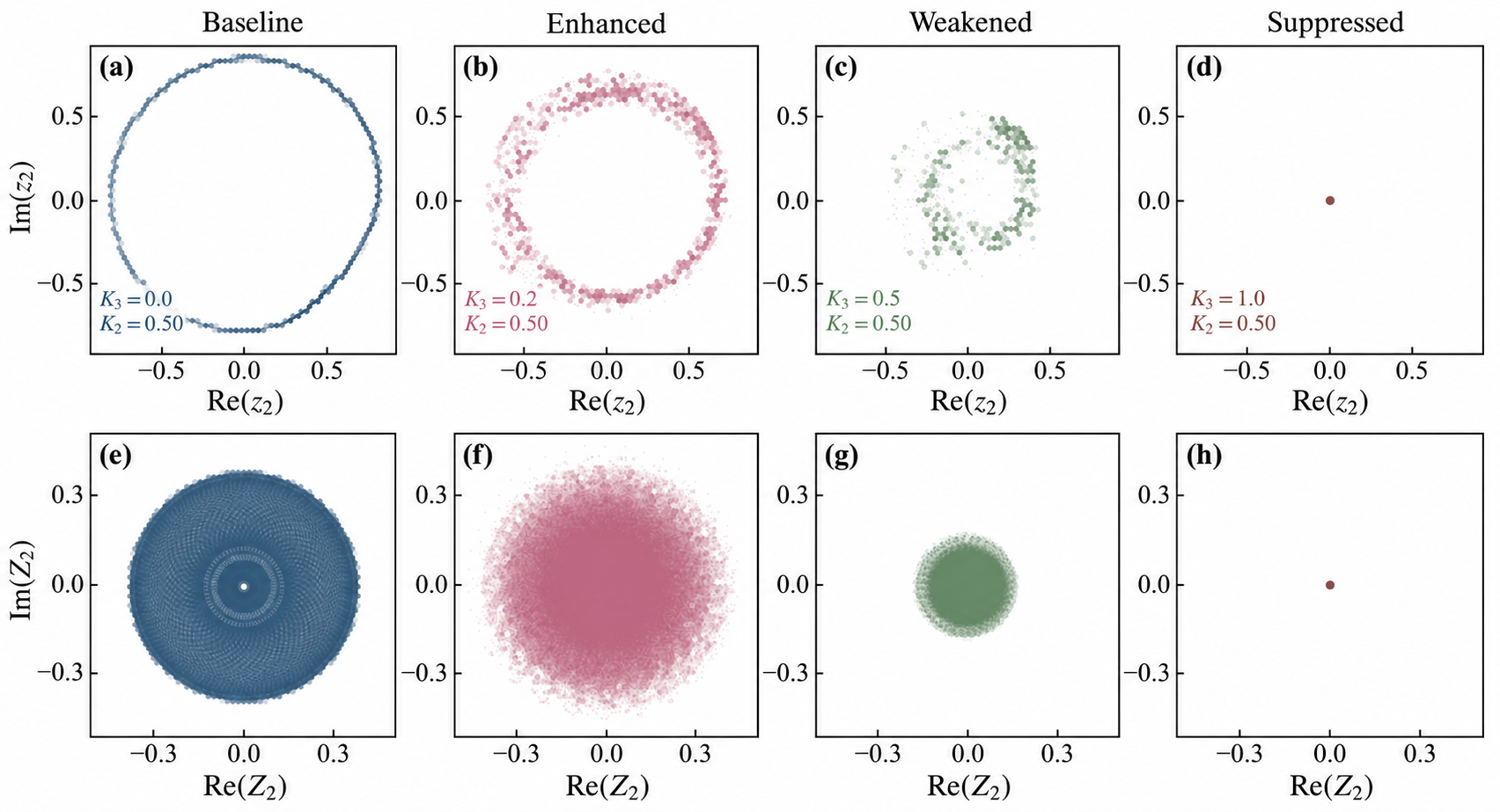}
\caption{
Attractor-geometry diagnostics of the preexisting chaotic branch at fixed \(K_2=0.50\) for representative \(K_3\). 
Columns show the baseline pairwise case, enhanced, weakened, and suppressed responses as \(K_3\) increases. 
(a)--(d) Poincar\'e sections of \(z_2\) sampled at upward crossings of \(\mathrm{Im}(z_1)=0\). 
(e)--(h) Corresponding trajectories in the collective plane \((\mathrm{Re}\,Z_2,\mathrm{Im}\,Z_2)\), with \(Z_2=N^{-1}\sum_{j=1}^{N}z_j^2\). 
The contraction and final collapse of these sets show how triadic coupling reshapes and eventually suppresses the chaotic branch.
}
\label{fig:geometry}
\end{figure*}

The geometry confirms this interpretation. In Fig.~\ref{fig:geometry}, weak $K_3$ produces a visibly thicker radial spread in the Poincar\'e section and a broader excursion in the collective plane, indicating that the branch explores a larger amplitude-supported region of phase space. At intermediate $K_3$ the same support contracts, and at strong $K_3$ the broad set collapses toward a regular point. The attractor is thus reconstructed rather than replaced. What is seen is a gradual compression of the same irregular geometry, followed eventually by its disappearance.

This geometric reading matters because it ties the window diagnostics back to the underlying state-space object. A narrowing of the chaotic support is not just the loss of positive Lyapunov points in a scan; it reflects the shrinking parameter range over which the amplitude-supported irregular geometry can persist. Likewise, the decay of the peak Lyapunov response is not merely a smaller number; it signals that even where the branch survives, its local stretching has weakened. The reconstruction language is therefore anchored simultaneously in parameter space, in tangent-space growth, and in attractor geometry.

To formalize that reconstruction, we define the connected chaotic window as the largest interval satisfying $\lambda_{\max}(K_2,K_3)>\lambda_c$ with $\lambda_c=0.05$. Its left and right edges are denoted $K_{2,\mathrm{on}}$ and $K_{2,\mathrm{off}}$, and the corresponding chaotic weight, equivalently the chaotic window width, is
\begin{equation}
W_{\mathrm{ch}}(K_3)=K_{2,\mathrm{off}}(K_3)-K_{2,\mathrm{on}}(K_3).
\label{eq:width_def}
\end{equation}
The strongest local instability inside that window is
\begin{equation}
\lambda_{\mathrm{peak}}(K_3)=\max_{K_2}\lambda_{\max}(K_2,K_3),
\label{eq:peak_lambda}
\end{equation}
and the onset sharpness is
\begin{equation}
S(K_3)=
\left.\frac{\dd \lambda_{\max}}{\dd K_2}\right|_{K_{2,\mathrm{on}}},
\label{eq:slope_def}
\end{equation}
with the derivative estimated from a local linear fit near onset. Eq.~(\ref{eq:width_def}) defines the chaotic weight, Eq.~(\ref{eq:peak_lambda}) defines the peak instability, and Eq.~(\ref{eq:slope_def}) defines the onset sharpness.

These diagnostics are intentionally formalized because they answer different questions about the same branch. $K_{2,\mathrm{on}}$ records where chaoticity first appears along the $K_2$ scan. $W_{\mathrm{ch}}$ measures how much parameter space remains inside the connected chaotic window. $\lambda_{\mathrm{peak}}$ measures the strongest local stretching attained within that window. $S(K_3)$ quantifies how abruptly the branch destabilizes at onset. None of these quantities is redundant with the others, and there is no reason for them to share a common dependence on $K_3$.

The onset sharpness is especially useful in this context because it discriminates between a branch that destabilizes abruptly and one that leaks into chaos more gradually. An increase of $S(K_3)$ means that the instability, once triggered, grows more steeply with $K_2$ near the left edge of the window. A later decrease of $S(K_3)$ therefore signals not only that the window is shrinking, but also that the remaining instability is becoming dynamically less abrupt. This additional diagnostic helps separate geometric weakening of the branch from a mere horizontal shift of its onset.

$W_{\mathrm{ch}}$ and $\lambda_{\mathrm{peak}}$ are therefore not equivalent observables. The former measures chaotic support, whereas the latter measures peak instability. A system can enter chaos early but sustain that chaos only over a narrow interval, or it can maintain a substantial chaotic interval while the strongest local stretching has already weakened. For the same reason a monotonic drift of $K_{2,\mathrm{on}}$ is not inconsistent with a nonmonotonic $W_{\mathrm{ch}}$, and neither quantity uniquely determines $S(K_3)$. The response to higher-order coupling is multidimensional even though it acts on a single preexisting branch.

\begin{figure}[t]
\centering
\includegraphics[width=1\linewidth]{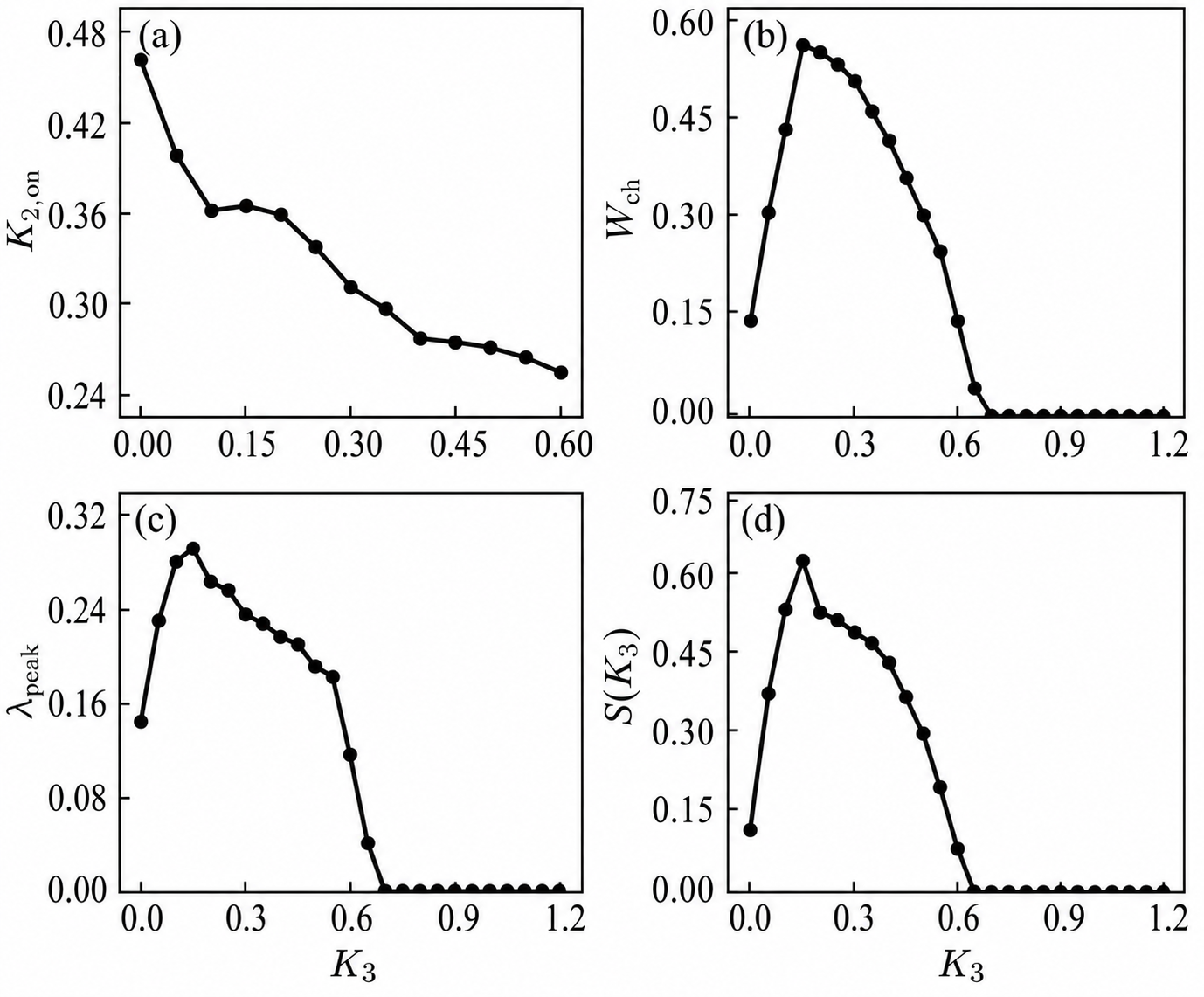}
\caption{
Formal diagnostics of how triadic coupling reshapes the reconstructed connected chaotic window. 
(a) Onset location \(K_{2,\mathrm{on}}\). 
(b) Chaotic-window width \(W_{\mathrm{ch}}\). 
(c) Peak largest Lyapunov exponent \(\lambda_{\mathrm{peak}}\). 
(d) Onset sharpness \(S(K_3)\). 
The four diagnostics show a nonmonotonic response: increasing \(K_3\) initially strengthens the preexisting chaotic branch but eventually narrows and suppresses it.
}
\label{fig:metrics}
\end{figure}

The formal diagnostics in Fig.~\ref{fig:metrics} show that the response of the preexisting chaotic branch to increasing $K_3$ is not uniform across diagnostics.
The onset location $K_{2,\mathrm{on}}$ shifts to smaller values approximately monotonically, indicating that chaoticity can emerge earlier in parameter space.
In contrast, the chaotic support $W_{\mathrm{ch}}$ exhibits a clear enhancement--suppression sequence, expanding at weak $K_3$ and shrinking at larger $K_3$.
The peak Lyapunov exponent $\lambda_{\mathrm{peak}}$ also varies nonmonotonically, but its dependence is not identical to that of $W_{\mathrm{ch}}$, reflecting the distinction between local stretching intensity and global chaotic support.
The onset sharpness $S(K_3)$ follows a similar enhancement--suppression pattern, increasing at intermediate $K_3$ and decreasing before the window disappears.

\section{Amplitude--Shear Mechanism}

The mechanism becomes explicit once the quartet is written in amplitude and phase variables. Writing
\begin{equation}
z_i=r_i e^{\ii\theta_i},
\qquad
Z=\rho e^{\ii\Phi},
\label{eq:polar_def}
\end{equation}
and multiplying Eq.~(\ref{eq:compact_model}) by $e^{-\ii\theta_i}$ gives
\begin{align}
\dot r_i+\ii r_i\dot\theta_i
={}&
\left(1+\ii\omega_i-(1+\ii c_i)r_i^2\right)r_i
\nonumber\\
&+K_2\left(\rho e^{\ii(\Phi-\theta_i)}-r_i\right)
\nonumber\\
&+K_3\left(\rho^2 e^{-\ii\theta_i}-r_i\right),
\label{eq:polar_intermediate}
\end{align}
which separates into the amplitude equation
\begin{align}
\dot r_i
={}&
(1-r_i^2)r_i
+K_2\!\left[\rho\cos(\Phi-\theta_i)-r_i\right]
\nonumber\\
&+K_3\!\left[\rho^2\cos\theta_i-r_i\right],
\label{eq:amp_eq}
\end{align}
and the phase equation
\begin{align}
\dot\theta_i
={}&
\omega_i-c_i r_i^2
+K_2\frac{\rho}{r_i}\sin(\Phi-\theta_i)
\nonumber\\
&-K_3\frac{\rho^2}{r_i}\sin\theta_i.
\label{eq:phase_eq}
\end{align}
The unique conversion channel appears in Eq.~(\ref{eq:phase_eq}) through the nonisochronic term $-c_i r_i^2$. If this term is removed, amplitude disorder can no longer generate shear.

This point is more restrictive than it may first appear. The pairwise and three-body terms in Eq.~(\ref{eq:phase_eq}) can redistribute phase differences directly, but by themselves they do not turn radial disorder into differential angular velocities. That conversion requires the amplitude-dependent frequency shift. In that sense nonisochronicity is not a secondary perturbation layered on top of the collective dynamics; it is the local route by which amplitude heterogeneity becomes phase-speed heterogeneity. This is why the later $k=0$ comparison has the status of a control experiment rather than a routine parameter check.

The instantaneous angular velocities can therefore be decomposed as
\begin{equation}
\Omega_i \equiv \dot\theta_i = \omega_i + \Omega_i^{\mathrm{sh}} + \Omega_i^{\mathrm{cpl}},
\end{equation}
with $\Omega_i^{\mathrm{sh}}=-c_i r_i^2=-k\omega_i r_i^2$ and $\Omega_i^{\mathrm{cpl}}=K_2\frac{\rho}{r_i}\sin(\Phi-\theta_i)-K_3\frac{\rho^2}{r_i}\sin\theta_i$. The shear term is therefore amplitude generated and frequency weighted. This is why $\sigma_\Omega$ is not merely a frequency statistic: it directly records the differential angular velocities that can drive shear-induced stretching. By contrast, $R$ remains only a coarse observable.

To identify the conversion step $\sigma_r \rightarrow \sigma_\Omega$, we write
\begin{equation}
\omega_i=\bar\omega+\delta\omega_i,
\qquad
r_i=\bar r+\delta r_i,
\qquad
\Omega_i=\bar\Omega+\delta\Omega_i,
\label{eq:fluct_defs}
\end{equation}
with zero-mean fluctuations. Using Eq.~(\ref{eq:phase_eq}) then gives
\begin{align}
\Omega_i
={}&
\bar\omega+\delta\omega_i
-k(\bar\omega+\delta\omega_i)(\bar r+\delta r_i)^2
+\Omega_i^{\mathrm{cpl}}
\nonumber\\
={}&
\bar\omega+\delta\omega_i
-k(\bar\omega+\delta\omega_i)
\left(
\bar r^2+2\bar r\,\delta r_i+(\delta r_i)^2
\right)
\nonumber\\
&+\Omega_i^{\mathrm{cpl}},
\label{eq:omega_expand}
\end{align}
and subtracting the oscillator mean yields
\begin{align}
\delta\Omega_i
={}&
(1-k\bar r^2)\delta\omega_i
-2k\bar\omega \bar r\,\delta r_i
\nonumber\\
&-2k\bar r\!\left(\delta\omega_i\delta r_i-C_{\omega r}\right)
+\delta\Gamma_i
+\mathcal R_i,
\label{eq:delta_omega}
\end{align}
where
\begin{align}
C_{\omega r}
={}&
\frac{1}{N}\sum_{j=1}^{N}\delta\omega_j\delta r_j,
\label{eq:cov_wr}
\\
\delta\Gamma_i
={}&
\Omega_i^{\mathrm{cpl}}-\overline{\Omega^{\mathrm{cpl}}},
\label{eq:delta_gamma}
\\
\mathcal R_i
={}&
-k\left[
(\bar\omega+\delta\omega_i)(\delta r_i)^2
-\overline{(\bar\omega+\delta\omega)(\delta r)^2}
\right].
\label{eq:remainder_i}
\end{align}
The fluctuation definitions in Eq.~(\ref{eq:fluct_defs}) therefore lead to the expansion in Eq.~(\ref{eq:omega_expand}) and the reduced fluctuation form in Eq.~(\ref{eq:delta_omega}), with the covariance and remainder terms defined in Eqs.~(\ref{eq:cov_wr})--(\ref{eq:remainder_i}). The point of the fluctuation expansion is not formal algebra alone. It identifies the leading conversion term, which is proportional to $\delta r_i$.

\begin{figure}[t]
\centering
\includegraphics[width=1\linewidth]{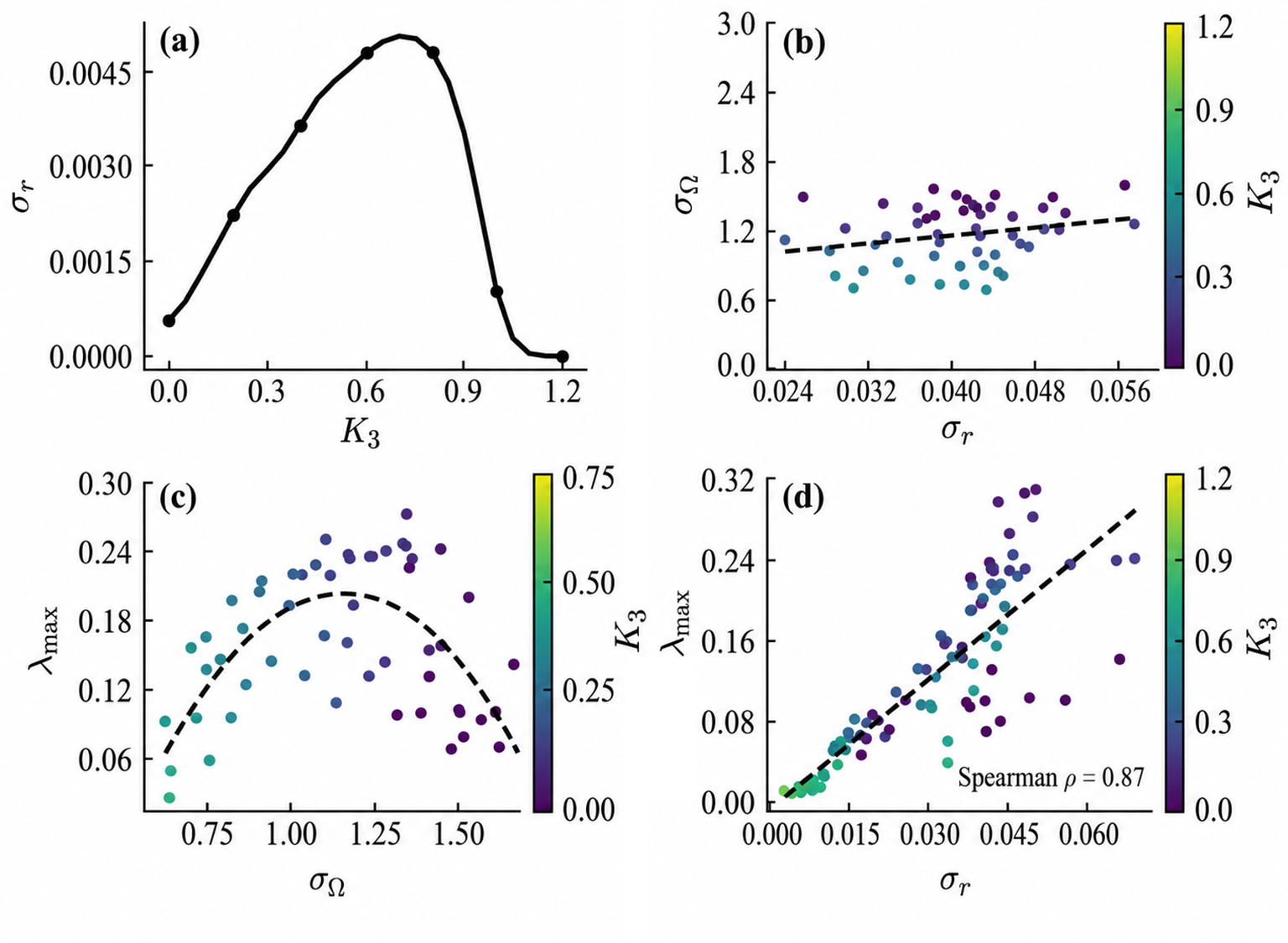}
\caption{
Mechanism closure along the amplitude--shear pathway. (a) Triadic coupling \(K_3\) regulates the amplitude dispersion \(\sigma_r\). (b) The amplitude dispersion is converted into effective-frequency shear \(\sigma_\Omega\) through nonisochronicity. (c) The largest Lyapunov exponent \(\lambda_{\max}\) is organized by the shear measure \(\sigma_\Omega\), showing that the instability response is more directly controlled by shear than by \(K_3\) itself. (d) When the same Lyapunov data are replotted against the upstream amplitude variance \(\sigma_r^2\), the responses follow an approximate common trend. This collapse indicates that higher-order coupling affects chaos mainly through amplitude-mediated shear. 
}
\label{fig:mechanism}
\end{figure}

Squaring Eq.~(\ref{eq:delta_omega}), averaging over oscillators, and retaining the leading contributions gives
\begin{align}
\sigma_\Omega^2
={}&
(1-k\bar r^2)^2\sigma_\omega^2
+4k^2\bar\omega^2\bar r^2\,\sigma_r^2
\nonumber\\
&+4k\bar\omega\bar r(1-k\bar r^2)\,C_{\omega r}
+\mathcal R_\Gamma
+\mathcal O(\sigma_r^3),
\label{eq:sigma_relation}
\end{align}
where $\sigma_\omega^2=\frac{1}{N}\sum_i\delta\omega_i^2$ is fixed by the imposed mismatch. Equation~(\ref{eq:sigma_relation}) shows that the dominant $K_3$ dependence reaches $\sigma_\Omega$ indirectly through $\sigma_r^2$. Higher-order coupling does not directly set the shear scale; instead, it first regulates amplitude dispersion and only then enters the angular-velocity spread. This is also why a direct plot of $\lambda_{\max}$ against $K_3$ is mechanistically incomplete: once the data are re-expressed in terms of $\sigma_\Omega$, much of the explicit $K_3$ dependence should disappear because the leading conversion step has already been absorbed. This expectation underlies the reduced collapse in Fig.~\ref{fig:mechanism}.

A finite $\sigma_\Omega$ implies differential angular velocities, and differential angular velocities imply local shear. Nearby states are then advected at unequal angular rates, stretched apart, reinjected by the bounded collective dynamics, and folded back into the accessible region of phase space. In that sense the argument is dynamical-systems based rather than correlational: shear supplies the differential rotation needed for stretching, while the bounded flow provides reinjection and folding. This is why $\sigma_\Omega$ is the controlling variable for the instability response, whereas a coherence measure can only register partial alignment.

The upstream step $K_3 \rightarrow \sigma_r$ follows from the amplitude sector. Expanding the local Stuart-Landau term around the instantaneous mean amplitude gives
\begin{equation}
(1-r_i^2)r_i
=(1-\bar r^2)\bar r-(3\bar r^2-1)\delta r_i+\mathcal O(\delta r_i^2),
\label{eq:local_expand}
\end{equation}
and subtracting the oscillator average of Eq.~(\ref{eq:amp_eq}) yields
\begin{align}
\dot{\delta r_i}
={}&
-\mu_r(K_2,K_3)\,\delta r_i
+K_2\rho\,\chi_i
+K_3\rho^2\,\eta_i
\nonumber\\
&+\mathcal N_i,
\label{eq:amp_dev}
\end{align}
with
\begin{align}
\mu_r(K_2,K_3)
={}&
3\bar r^2+K_2+K_3-1,
\label{eq:mu_r}
\\
\chi_i
={}&
\cos(\Phi-\theta_i)-\overline{\cos(\Phi-\theta)},
\label{eq:chi_def}
\\
\eta_i
={}&
\cos\theta_i-\overline{\cos\theta}.
\label{eq:eta_def}
\end{align}
and nonlinear remainder $\mathcal N_i$. The local amplitude expansion in Eq.~(\ref{eq:local_expand}) thus yields the deviation dynamics in Eq.~(\ref{eq:amp_dev}), with the restoring rate and geometric coefficients defined in Eqs.~(\ref{eq:mu_r})--(\ref{eq:eta_def}). The three-body term therefore acts in two opposing ways: it injects amplitude heterogeneity through $K_3\rho^2\eta_i$, but it also increases the amplitude-restoring rate $\mu_r$.

Multiplying Eq.~(\ref{eq:amp_dev}) by $\delta r_i$ and summing over oscillators gives
\begin{align}
\frac{1}{2}\frac{\dd}{\dd t}\sum_{i=1}^{N}\delta r_i^2
={}&
-\mu_r\sum_{i=1}^{N}\delta r_i^2
+K_2\rho\sum_{i=1}^{N}\delta r_i\chi_i
\nonumber\\
&+K_3\rho^2\sum_{i=1}^{N}\delta r_i\eta_i
+\sum_{i=1}^{N}\delta r_i \mathcal N_i,
\label{eq:amp_balance}
\end{align}
which makes the nonmonotonicity transparent. Weak $K_3$ first enlarges $\sigma_r$ because heterogeneity injection grows before restoring dominates. Strong $K_3$ then suppresses $\sigma_r$ because the amplitude-restoring rate becomes large enough to collapse the amplitude--shear feedback loop.

The intermediate regime follows naturally from the same balance. Once the restoring contribution proportional to $\mu_r$ has become comparable to the heterogeneity injection term, the system can still destabilize early in $K_2$ because the conversion channel remains active, yet the total amount of sustained amplitude disorder is already declining. In that regime the branch can onset sooner while carrying a smaller reservoir of radial fluctuations. This is exactly the situation in which onset location decouples from chaotic support and peak instability. The nonmonotonic reconstruction described in the previous section is therefore the observable footprint of a competition internal to the amplitude sector.

The mechanism is closed in Fig.~\ref{fig:mechanism}. Panel (a) validates the upstream regulation $K_3 \rightarrow \sigma_r$. Panel (b) validates the conversion step $\sigma_r \rightarrow \sigma_\Omega$ rather than a generic correlation. Panel (c) shows that $\lambda_{\max}$ collapses onto a one-dimensional reduced law when expressed against $\sigma_\Omega^2$, thereby recasting chaos as a shear-organized  response instead of a direct function of $K_3$. Panel (d) shows the same scan data replotted against the internal amplitude variance $\sigma_r^2$, where the apparent linear trend provides the upstream amplitude-level signature of the shear-organized Lyapunov response. The full chain is therefore closed as
\begin{equation}
K_3 \rightarrow \sigma_r \rightarrow \sigma_\Omega \rightarrow \lambda_{\max}.
\label{eq:chain}
\end{equation}

The master-curve collapse has a specific theoretical meaning. It does not claim exact universality of every residual term, nor does it say that higher-order coupling leaves no trace once the data are reparametrized. Rather, it shows that the leading-order dependence on $K_3$ has been absorbed into the upstream regulation of $\sigma_r$ and the subsequent generation of $\sigma_\Omega$. After that projection, the Lyapunov response becomes approximately one dimensional. This is the strongest evidence in the paper that chaos is not directly controlled by the higher-order parameter itself. By contrast, no comparable collapse is obtained from the coarse order parameter $R$, which is precisely why $R$ must remain a secondary diagnostic.

Equally important, the collapse explains how several apparently separate observations can coexist without contradiction. The onset may shift left, the width may increase and then decrease, and the peak instability may follow its own trend, yet all of these changes can still be consistent with a single shear coordinate. What varies from one diagnostic to another is how each quantity samples the reconstructed branch. What remains common is that the branch gains or loses instability only insofar as the accessible shear is amplified or suppressed.

\begin{figure*}[t]
    \centering
    \includegraphics[width=1\linewidth]{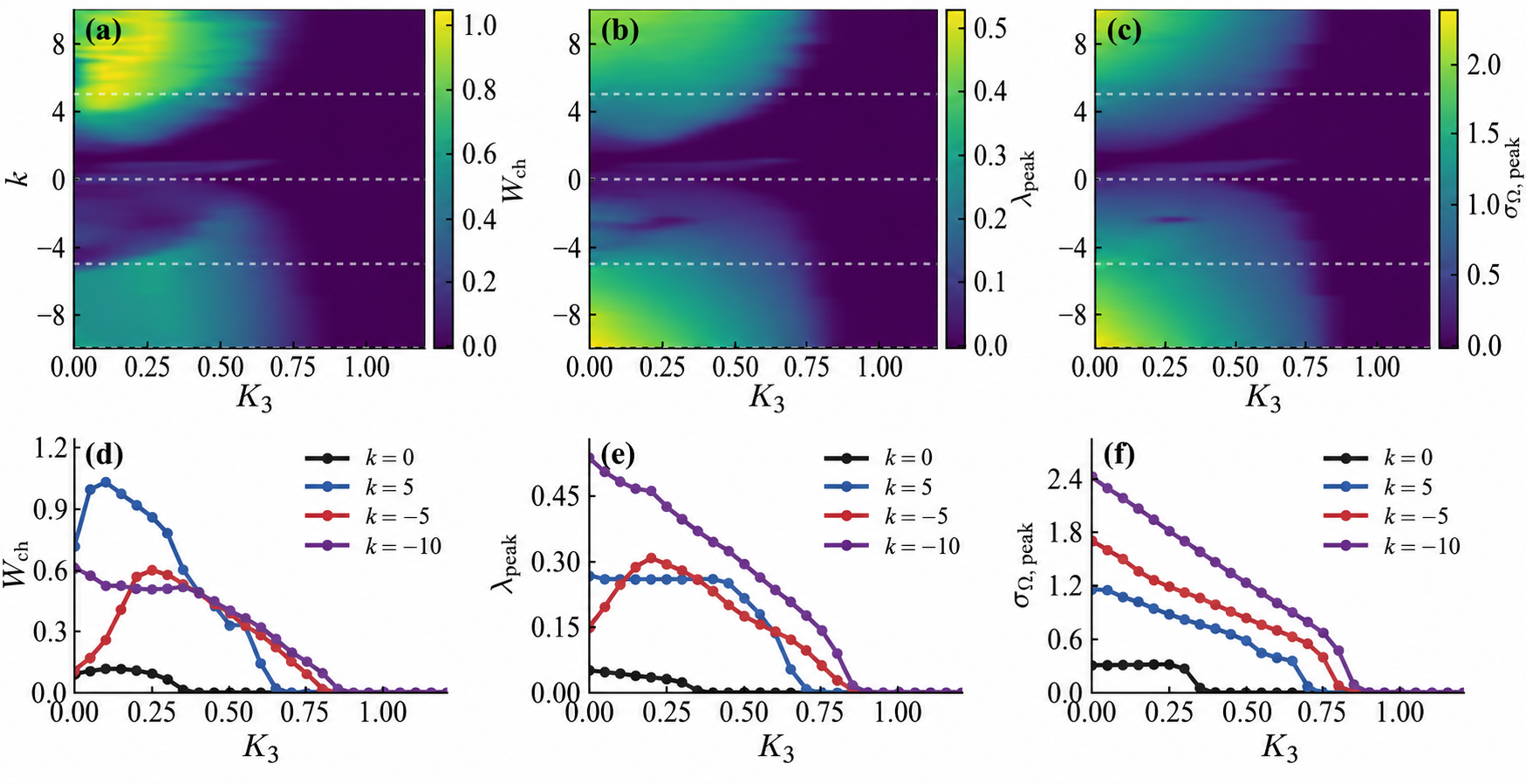}
    \caption{Mechanism validation in the \((K_3,k)\) plane. 
(a)--(c) Heat maps of the chaotic-window width \(W_{\mathrm{ch}}\), peak Lyapunov exponent \(\lambda_{\mathrm{peak}}\), and peak effective-frequency shear \(\sigma_{\Omega,\mathrm{peak}}\). 
(d)--(f) \(K_3\)-dependent slices for representative \(k\), with \(k=0\) serving as the control case without amplitude-to-phase conversion. 
The joint variation of \(W_{\mathrm{ch}}\), \(\lambda_{\mathrm{peak}}\), and \(\sigma_{\Omega,\mathrm{peak}}\) shows that the chaotic response is organized by the shear channel across the \((K_3,k)\) plane.}
\label{fig:k_scan}
\end{figure*}

The mechanism claim becomes substantially stronger only if it survives beyond the original one-parameter scan in $K_2$. To test this, both the higher-order coupling $K_3$ and the amplitude-to-phase conversion strength $k$ are varied, as shown in Fig.~\ref{fig:k_scan}.
Across the $(K_3,k)$ plane, regions of large chaotic support $W_{\mathrm{ch}}$, large peak instability $\lambda_{\mathrm{peak}}$, and large peak effective-frequency shear $\sigma_{\Omega,\mathrm{peak}}$ are broadly co-located. 
This correspondence is structural rather than pointwise: the different diagnostics quantify distinct aspects of the same branch and are therefore not expected to coincide exactly. 
The role of the two-parameter scan is thus not to map a phase diagram, but to test whether the amplitude--shear pathway persists over a broader domain.

The case $k=0$ provides a control in which the amplitude-to-phase conversion channel is removed, since $c_i = k \omega_i = 0$. 
In this limit, amplitude heterogeneity can still be generated in the radial sector, but it can no longer be converted into effective-frequency shear. 
As a result, $\sigma_\Omega$ is strongly suppressed across the $K_3$ scan. 
A weak chaotic window may nevertheless persist due to the preexisting pairwise baseline dynamics. 
In this regime, the characteristic enhancement--suppression structure disappears: both $\lambda_{\mathrm{peak}}$ and $W_{\mathrm{ch}}$ decrease more smoothly with increasing $K_3$, without exhibiting the pronounced nonmonotonic reconstruction observed at finite $k$. 
This comparison shows that shear is not required for chaos to exist, but it is required for the higher-order reconstruction of chaos.

Varying $k$ away from zero then tunes the efficiency of the same conversion pathway. 
Larger $|k|$ increases the sensitivity of angular velocities to amplitude fluctuations and enhances the resulting shear, while smaller $|k|$ weakens this transduction. 
The two-parameter scans therefore probe the two middle links of the causal chain, $K_3 \rightarrow \sigma_r$ and $\sigma_r \rightarrow \sigma_\Omega$, rather than introducing an independent control mechanism. 
The observed co-location of the response sectors is consistent with shear acting as the controlling variable of the instability, rather than coherence alone.

The broader scans also clarify why the window diagnostics must remain distinct. 
$W_{\mathrm{ch}}$ measures the extent of chaotic support in parameter space, whereas $\lambda_{\mathrm{peak}}$ measures the strongest local stretching rate attained within that support. 
Accordingly, $W_{\mathrm{ch}}$ can retain a strongly nonmonotonic enhancement--suppression profile even when $\lambda_{\mathrm{peak}}$ varies more smoothly with increasing $K_3$. 
This separation is not a numerical inconsistency, but the expected signature of a reconstructed chaotic branch whose support and peak instability respond differently to the same upstream amplitude regulation.

\begin{figure}[t]
\centering
\includegraphics[width=1\linewidth]{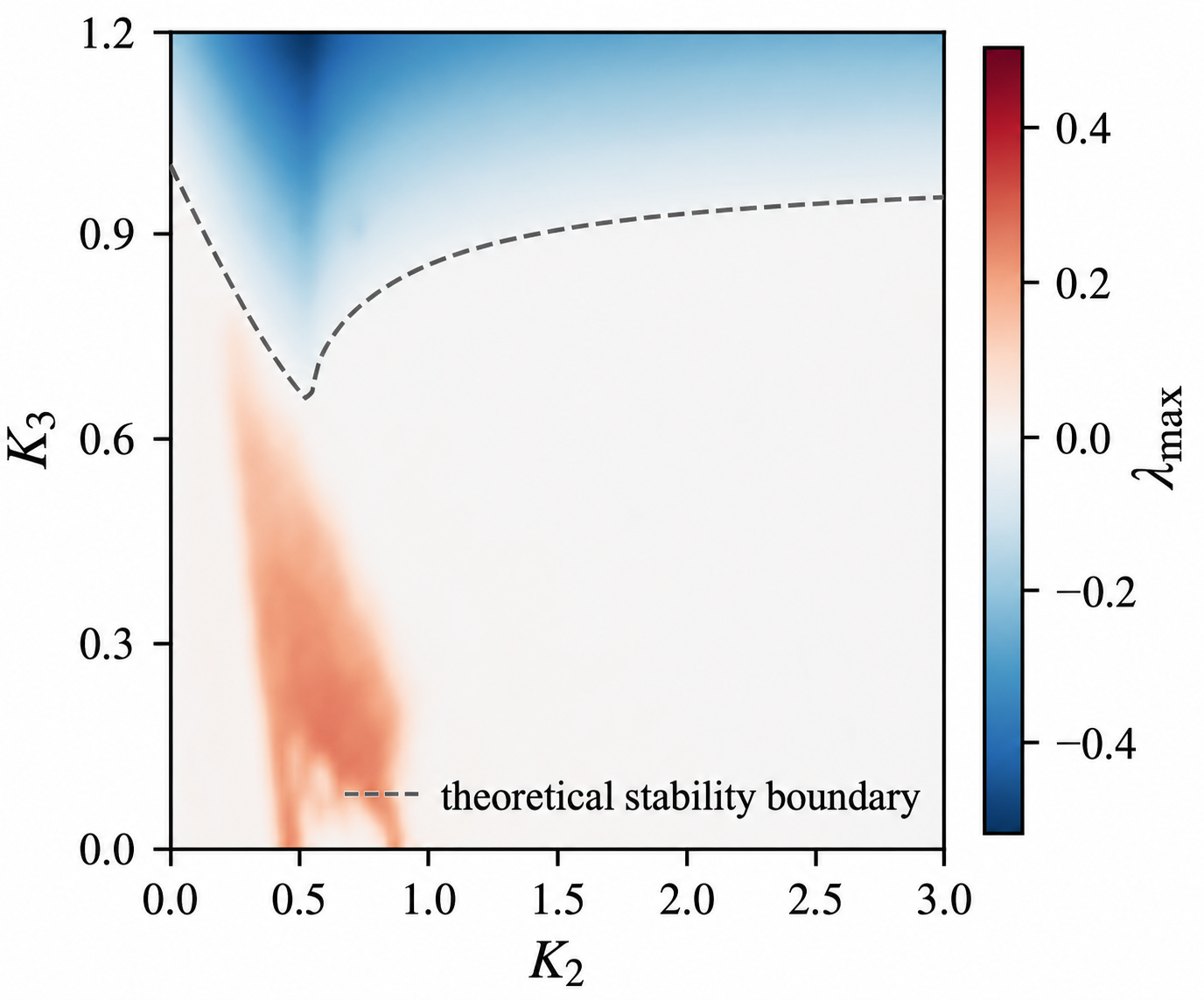}
\caption{
Lyapunov map and analytical stability boundary in the \((K_2,K_3)\) plane. 
Colors denote the numerically computed largest Lyapunov exponent \(\lambda_{\max}\), with positive values marking chaos. 
The dashed curve is the theoretical stability boundary of the phase-fixed regular state, \(\Lambda_{\mathrm{reg}}=0\), defined in Eq.~\eqref{eq:lambda_reg}. 
Its alignment with the upper edge of the chaotic region indicates that chaos disappears at large \(K_3\) through stabilization of the regular branch.
}
\label{fig:stable}
\end{figure}

The same interpretation admits a linear-stability projection in the full $(K_2,K_3)$ plane. The dashed curve in Fig.~\ref{fig:stable} is not a fit to the Lyapunov map and not an independent appendix result. Because the triadic term $K_3(|Z|^2-z_i)$ fixes an absolute phase reference, the regular branch is continued directly in the laboratory frame rather than as a neutrally rotating branch. We therefore write the phase-pinned regular state as
\begin{equation}
z_i(t)=u_i^\ast,
\qquad
u_i^\ast=r_i^\ast e^{\ii\phi_i^\ast},
\qquad
U^\ast=\frac{1}{N}\sum_{j=1}^{N}u_j^\ast .
\label{eq:regular_branch_ansatz}
\end{equation}
The branch satisfies
\begin{align}
0
={}&
\left[
1+\ii\omega_i
-(1+\ii c_i)|u_i^\ast|^2
\right]u_i^\ast
\nonumber\\
&+K_2(U^\ast-u_i^\ast)
+K_3(|U^\ast|^2-u_i^\ast).
\label{eq:regular_branch_condition}
\end{align}
This condition is the fixed-point equation of the laboratory-frame dynamics. It is the appropriate continuation problem here because the higher-order term removes the neutral global phase direction that would otherwise be present in a purely rotation-equivariant Stuart--Landau network.

Linearizing Eq.~\ref{eq:amp_eq} and Eq.~\ref{eq:phase_eq} around this
phase-pinned branch gives
\begin{equation}
\frac{\dd}{\dd t}
\begin{pmatrix}
\mathbf a\\
\bm\eta
\end{pmatrix}
=
\mathcal J_{\mathrm{reg}}
\begin{pmatrix}
\mathbf a\\
\bm\eta
\end{pmatrix},
\qquad
\mathcal J_{\mathrm{reg}}
=
\begin{pmatrix}
A_\ast & B_\ast\\
C_\ast & D_\ast
\end{pmatrix},
\label{eq:jreg_block}
\end{equation}
where the perturbations are defined by
$r_i=r_i^\ast+a_i$ and $\theta_i=\phi_i^\ast+\eta_i$. The block entries
are
\begin{align}
(A_\ast)_{ij}
={}&
\left.\frac{\partial \dot r_i}{\partial r_j}\right|_\ast,
\qquad
(B_\ast)_{ij}
=
\left.\frac{\partial \dot r_i}{\partial \theta_j}\right|_\ast,
\label{eq:block_ab}
\\
(C_\ast)_{ij}
={}&
\left.\frac{\partial \dot\theta_i}{\partial r_j}\right|_\ast,
\qquad
(D_\ast)_{ij}
=
\left.\frac{\partial \dot\theta_i}{\partial \theta_j}\right|_\ast .
\label{eq:block_cd}
\end{align}
The largest real part of the spectrum,
\begin{equation}
\Lambda_{\mathrm{reg}}(K_2,K_3)
=
\max \Re \operatorname{eig}\!\left(\mathcal J_{\mathrm{reg}}\right),
\label{eq:lambda_reg}
\end{equation}
defines the stability boundary $\Lambda_{\mathrm{reg}}=0$ plotted in Fig.~\ref{fig:stable}. The purpose of this construction is not to
reproduce every feature of the chaotic wedge, but to test whether the same amplitude-mediated feedback loop identified in the nonlinear mechanism is already visible in the linearized regular branch.

Because the regular branch is amplitude locked, the amplitude block can be adiabatically eliminated when $A_\ast$ remains stable:
\begin{equation}
\mathbf a\simeq -A_\ast^{-1}B_\ast \bm\eta .
\label{eq:adiabatic_amp}
\end{equation}
This yields the reduced phase operator
\begin{equation}
\dot{\bm\eta}
=
\left(
D_\ast-C_\ast A_\ast^{-1}B_\ast
\right)\bm\eta .
\label{eq:reduced_phase_operator}
\end{equation}
The Schur-complement term $C_\ast A_\ast^{-1}B_\ast$ is therefore the
linearized amplitude-to-phase feedback channel. Near a nearly uniform regular branch, the amplitude block is dominated by the local restoring rate,
\begin{equation}
A_\ast \approx -\mu_r^\ast I,
\qquad
\mu_r^\ast \approx 3\bar r_\ast^2+K_2+K_3-1 .
\label{eq:local_a_block}
\end{equation}
Equation~\ref{eq:local_a_block} explains why the feedback channel weakens at large $K_3$: the effective amplitude-restoring rate
$\mu_r^\ast$ increases, so amplitude perturbations are damped before they
can feed phase shear efficiently. Thus the stability projection in
Fig.~\ref{fig:stable} is the laboratory-frame linear counterpart of the nonlinear amplitude--shear mechanism summarized in Eq.~\eqref{eq:chain}.

This linear perspective is the parameter-plane counterpart of Fig.~\ref{fig:mechanism}. The nonlinear mechanism states that strong three-body coupling collapses the amplitude--shear loop by suppressing amplitude disorder. The linearized continuation says the same thing in a different language: as $K_3$ increases, the regular amplitude-locked branch becomes harder to destabilize through amplitude-mediated phase feedback. The termination of the chaotic wedge is therefore consistent with the loss of the same feedback loop, not with the emergence of a separate organizing principle.

That correspondence is important for the logic of the manuscript. If the dashed curve merely shadowed the numerical wedge without sharing its mechanism, the analytical extension would add little beyond a visual guide. Its value is that the same amplitude-restoring trend that explains the nonlinear suppression of $\sigma_r$ also weakens the linearized feedback operator. Fig.~\ref{fig:stable} is therefore best read as a structural projection of the amplitude--shear mechanism into the full parameter plane, not as an independent result competing with Fig.~\ref{fig:mechanism}.

\begin{figure}[t]
    \centering
    \includegraphics[width=1\linewidth]{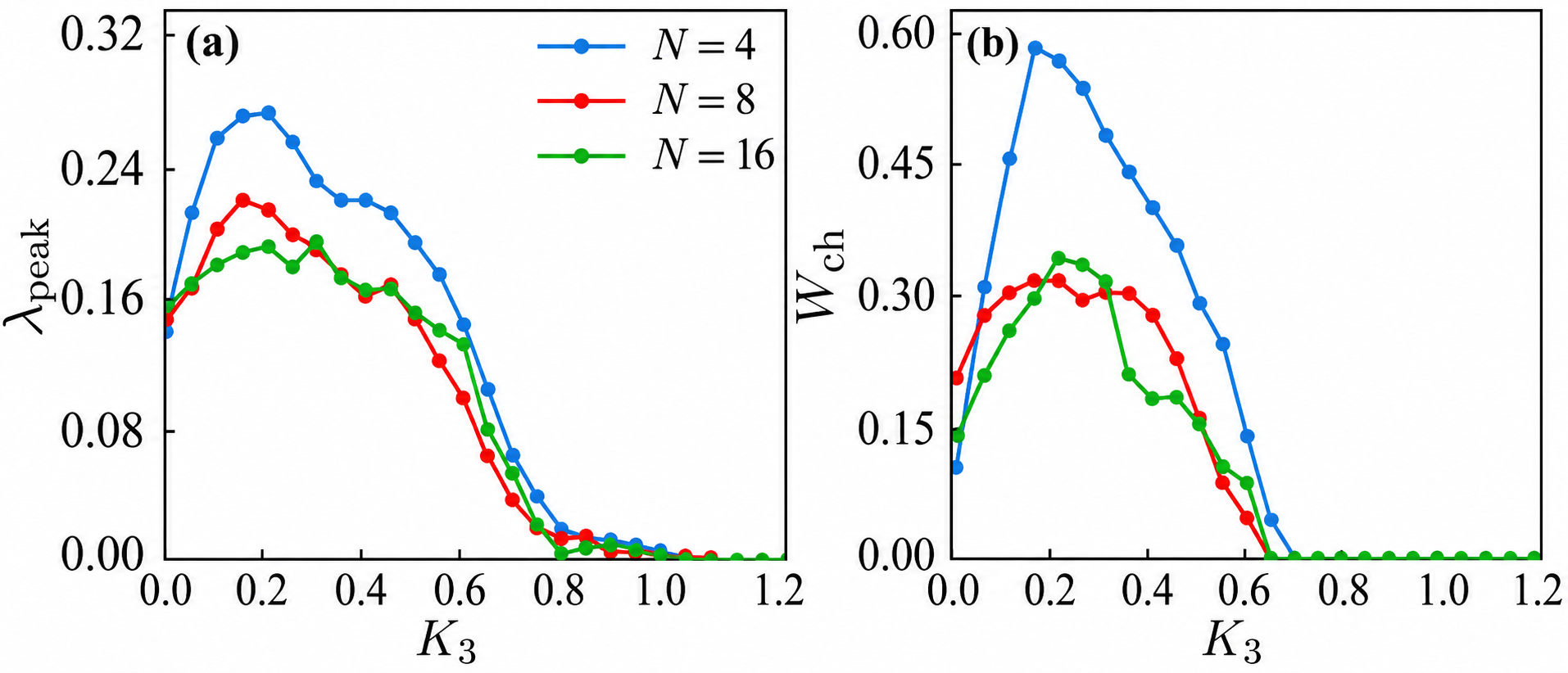}
    \caption{
Finite-size extension of the enhancement--suppression sequence. 
(a) Peak largest Lyapunov exponent \(\lambda_{\mathrm{peak}}\) and 
(b) chaotic-window width \(W_{\mathrm{ch}}\) versus \(K_3\) for fully connected systems with \(N=4,8\), and \(16\). 
The same nonmonotonic trend persists with increasing system size, showing that weak triadic coupling enhances the chaotic branch while stronger triadic coupling suppresses it.
}
    \label{fig:Multi_N}
\end{figure}

The final test is whether the quartet is merely a special numerical case or a genuine minimal mechanism carrier. Fig.~\ref{fig:Multi_N} addresses that question by extending the same diagnostics to fully connected systems with $N=8$ and $N=16$, keeping the mismatch structure and the frequency-dependent nonisochronicity $c_i=k\omega_i$. The goal is not to claim strict universality in every quantitative detail. It is to determine whether the same indirect control law remains visible once the system size increases.

The answer is positive at the level of mechanism. The $K_3$-induced enhancement--suppression sequence persists at larger $N$: weak higher-order coupling still increases the peak instability and the chaotic window width, while strong higher-order coupling still suppresses both. At the same time, increasing system size reduces the typical fluctuation amplitude, weakens the achievable effective-frequency shear, lowers $\lambda_{\mathrm{peak}}$, and narrows $W_{\mathrm{ch}}$. Larger ensembles therefore do not change the identity of the controlling variable, but they do change how strongly it can be excited.

The finite-size extension should thus be read in a restrained way. The mechanism is robust, but its strength is size modulated. The quartet is valuable precisely because it resolves the full pathway cleanly; larger systems preserve the same hierarchy $K_3 \rightarrow \sigma_r \rightarrow \sigma_\Omega \rightarrow \lambda_{\max}$ while averaging away part of the fluctuation reservoir on which that pathway relies. In that sense the quartet is not an accidental special case, and it is not a surrogate thermodynamic limit either. It is the smallest setting in which the amplitude--shear pathway can be read with minimal ambiguity.

Seen from this perspective, the finite-size results also clarify what should and should not be expected in larger ensembles. One should not expect the same numerical thresholds, the same peak values, or the same window widths to persist unchanged, because collective averaging alters the accessible fluctuation scale. One should expect the causal hierarchy to persist as long as amplitude heterogeneity remains dynamically relevant and nonisochronicity continues to convert it into shear. That is the level at which the mechanism is robust.

This distinction between mechanistic robustness and quantitative identity is especially important for future extensions. If larger or more heterogeneous systems display broader spectra of collective states, one should not judge the relevance of the present quartet solely by whether every threshold shifts in the same way. The more meaningful test is whether higher-order feedback still acts primarily through the amplitude field and whether the resulting shear remains the decisive instability coordinate. The finite-size extension supports that more careful interpretation.

\section{Discussion}

The main conceptual shift of this work is that higher-order control of chaos should not be framed primarily as a problem of coherence. In the present system, coherence is informative but not organizing. What determines whether the irregular branch survives is whether amplitude heterogeneity remains dynamically available and whether it is converted into differential angular velocities. Once that internal route weakens, chaos disappears even though partial incoherence may still be visible. The route to chaos is therefore organized by an internal amplitude--shear process rather than by a single synchronization coordinate.

Accordingly, the three-body channel should be interpreted less as an extra synchronizing or desynchronizing force and more as a regulator of the amplitude field. This changes the meaning of higher-order coupling. Weak $K_3$ does not simply ``enhance chaos,'' nor does strong $K_3$ simply ``suppress chaos'' in a direct sense. Instead, the many-body term first reshapes the availability of amplitude disorder; only afterward does that change appear in the shear sector and, finally, in the instability output. Higher-order interactions thus act on a hidden internal degree of freedom and only indirectly on the observable chaotic response.

This viewpoint also changes what one should measure. In nonisochronous amplitude oscillators, a global order parameter can be too coarse to diagnose why an irregular state is gained or lost. A state may appear only moderately coherent and yet be dynamically regular because the amplitude field has already been strongly damped. Conversely, a partially coherent state may remain chaotic if internal amplitude fluctuations still sustain sufficient angular-velocity dispersion. The more informative question is therefore not simply how aligned the oscillators are, but whether the system still supports the internal shear needed for stretching and reinjection.

The quartet matters because it isolates that causal structure with minimal ambiguity. Its value is not asymptotic but analytical: it is the smallest globally coupled setting in which the full pathway from higher-order regulation to instability can be resolved cleanly. The finite-size extension then suggests what is likely to persist beyond this minimal carrier. One should not expect the same thresholds or window widths to remain unchanged in sparse, noisy, or motif-rich systems, nor in systems with multiple irregular branches. One may nevertheless expect the causal hierarchy to be more robust than any particular number. In that sense, the present work changes the interpretation of higher-order control of chaos: higher-order interactions are best understood as regulators of hidden internal degrees of freedom whose consequences only later appear as instability.

\section{Conclusions}

In this work, we have identified an amplitude--shear mechanism that organizes the chaotic response of a nonisochronous Stuart--Landau quartet under higher-order coupling. 
Rather than creating chaos from a regular background, the three-body interaction reconstructs a preexisting chaotic branch by acting on its internal amplitude dynamics.

This reconstruction follows a structured pathway. 
Higher-order coupling first regulates amplitude heterogeneity, which is then converted into effective-frequency shear through frequency-dependent nonisochronicity. 
The resulting shear determines how instability is expressed along the branch, including its strength, extent, and nonmonotonic reconstruction.

The mechanism closes as $K_3 \rightarrow \sigma_r \rightarrow \sigma_\Omega \rightarrow \lambda_{\max}$, identifying shear as the organizing variable of the chaotic response. 
Consistently, removing the amplitude-to-phase conversion channel suppresses shear and eliminates the characteristic enhancement--suppression structure, while leaving a weak baseline chaotic window intact.

These results show that higher-order interactions act by regulating internal dynamical degrees of freedom rather than directly modifying coherence. 
In this sense, shear controls how instability is expressed along the branch.

\section*{CODE AVAILABILITY}

The source code required to reproduce the main results is available online as Ref.~\cite{routechaoscode}.

\bibliographystyle{apsrev4-2}
\bibliography{refs}

\end{document}